\documentclass[aps,prb,preprint,superscriptaddress,showpacs]{revtex4}
\usepackage{graphicx}
\usepackage{dcolumn}
\usepackage{bm}
\usepackage{amssymb}
\usepackage{color}

\begin{document}

\title{Pressure-Induced Structural Phase Transition in CeNi: X-ray and Neutron Scattering Studies and First-Principles Calculations}

\author{A.~Mirmelstein}
\affiliation{Department of Experimental Physics, Russian Federal Nuclear Center - E. I. Zababakhin Research Institute of Technical Physics (RFNC-VNIITF), Snezhinsk, 456770, Russia}
\author{A.~Podlesnyak}
\thanks{Corresponding author. Electronic address: podlesnyakaa@ornl.gov}
\author{António M. dos Santos}
\author{G.~Ehlers}
\affiliation{Quantum Condensed Matter Division, Oak Ridge National Laboratory, Oak Ridge, Tennessee 37831, USA}
\author{O.~Kerbel}
\author{V.~Matvienko}
\affiliation{Department of Experimental Physics, Russian Federal Nuclear Center - E. I. Zababakhin Research Institute of Technical Physics (RFNC-VNIITF), Snezhinsk, 456770, Russia}
\author{A. S.~Sefat}
\author{B.~Saparov}
\affiliation{Materials Science and Technology Division, Oak Ridge National Laboratory, Oak Ridge, TN 37831, USA}
\author{G. J.~Halder}
\affiliation{X-ray Science Division, Argonne National Laboratory, Argonne, IL 60439, USA}
\author{J. G.~Tobin}
\affiliation{Lawrence Livermore National Laboratory, CA 94550, USA}

\date{\today}

\begin{abstract}
The pressure-induced structural phase transition in the intermediate-valence compound CeNi has been investigated by X-ray and neutron powder diffraction techniques. For the first time it is shown that the structure of the pressure-induced CeNi phase (phases) can be described in terms of the $Pnma$ space group. Equations of state for CeNi on both sides of the phase transition are derived and an approximate $P-T$ phase diagram is suggested for $P < 8$~GPa and $T < 300$~K. The observed $Cmcm \rightarrow Pnma$ structural transition is analyzed using density functional theory (DFT) calculations, which successfully reproduce the ground state volume, the phase transition pressure, and the volume collapse associated with the phase transition.
\end{abstract}


\maketitle

\section{INTRODUCTION}

The 4$f$-metals, and their alloys and intermetallic compounds exhibit rich pressure-temperature-composition phase diagrams. Often, transitions between different phases are accompanied by a dramatic change in the material properties, including abrupt volume variation, and in particular the extent to which the $f$ electron localization changes under such phase transitions is of great scientific interest.

The transition between localized and itinerant behaviour is at the forefront of physics of strongly correlated systems. Such a transition occurs in the 5$f$ electron shell of the pure actinide metals - the 5$f$ electrons behave in a localized fashion in the heavy actinides but in a more delocalized manner in the light actinides - centering near Pu and Am. A quantitative and conclusive understanding of this phenomenon is still missing.\cite{Tobin} In the lanthanide series, where the 4$f$ orbitals are spatially less extended than the 5$f$ orbitals in the actinides, localized-itinerant transitions occur at the beginning (around Ce), in the middle (Sm, Eu) and at the end (Yb). Application of high pressure leads to a 4$f$ delocalization and, in some cases, to structural phase transformations accompanied by a volume discontinuity. The most famous example is cerium, which exhibits the isostructural $\gamma\rightarrow\alpha$ volume-collapse phase transition upon either cooling or application of external pressure. Several interpretations have been suggested to explain the origin of this phenomenon however its nature is still widely debated.\cite{Chakrabarti,Lanata1,Lanata2}
The intermediate-valence compound CeNi represents another interesting example of a system experiencing a pressure-induced structural instability.

CeNi has the CrB-type orthorhombic crystal structure (space group $Cmcm$) with the room-temperature crystal lattice parameters $a=3.77$~\AA, $b=10.46$~\AA, $c=4.37$~\AA.\cite{Tinney}
The orthorhombic $Cmcm$ crystal structure repeatedly appears in rare-earth and actinide metals under pressure such as $\alpha^{\prime}$-Ce, Pa, Nd, and Pr.
The $\alpha$-U metal also has this type of structure which is usually stable up to very high pressures.
Of course, at hundreds of GPa of pressures, closed-packed metal structures, such as fcc, hcp, and bcc, are once again favored due to large electrostatic repulsions ruling out the more open and lower symmetry structures.
CeNi shows a different behavior. Already at ambient pressure CeNi displays clear signatures of lattice instability upon cooling,\cite{Clementyev} but a structural transformation does not occur down to the lowest temperature.
In 1985 Gignoux and Voiron found a pressure-induced first-order structural phase transition in CeNi and determined its phase $P-T$ diagram up to $\sim$0.5~GPa and $\sim$150~K.\cite{Gignoux1,Gignoux2}
Later the CeNi phase diagram was extended up to $\sim$2~GPa and 300~K using neutron diffraction and magnetic measurements.\cite{Mirmelstein}
However, neither the space group nor atomic positions of the pressure-induced CeNi phase were established at the time.

Here we report the results of two diffraction studies performed to investigate the pressure-induced structural phase transition in CeNi. The first study was performed at room temperature using X-ray diffraction, and the second study employed neutron powder diffraction at 100~K.
In both the cases we observed a pressure-induced structural phase transition in CeNi accompanied by a volume jump.
The results obtained show that the structure of pressure-induced CeNi phase (phases) can be described with the $Pnma$ space group.
We show that density functional theory (DFT) calculations successfully reproduce the ground state volume,
the phase transition pressure, and the volume collapse associated with the phase transition.

\section{EXPERIMENTAL}

Stoichiometric amounts of elemental reactants, Ce (chunks, Ames) and Ni (Alfa) were loaded into alumina crucibles, which were baked at 800$^{\circ}$~C prior to use, and then placed inside quartz tubes.
Additional alumina crucibles containing Zr pieces were placed on top as potential oxygen sponges during sample synthesis process.
The quartz tubes were sealed under vacuum, and subsequently heated in box furnaces at 800$^{\circ}$~C for 6 hours, followed by 1$^{\circ}$~C/h cooling to 600$^{\circ}$~C.
The phase purity of the synthesized samples was verified using X-ray powder diffraction and magnetization measurements which showed that the desired CeNi phase ($Cmcm$ space group) had formed with a small amount of an impurity phase ($<2$\% of CeNi$_2$).
Magnetization measurements performed with a commercial Physical Properties Measurement System (Quantum Design) showed good agreement with the published data.\cite{Clementyev}

Room temperature high-pressure synchrotron  powder diffraction measurements were carried out at beamline 17-BM at the Advanced Photon Source using a 100 micron monochromated X-ray beam at a wavelength of $\lambda = 0.727750$~\AA.
Variable pressure diffraction data were collected \textit{in situ} using a Perkin Elmer amorphous-Si flat panel detector  centered on the X-ray beam.
The sample-to-detector distance was nominally set at 300~mm, yielding an available 2$\Theta$ scattering angle of 27.5~deg., corresponding to access of Bragg reflections with $d$-spacing as low as 1.52~\AA.
The diffractometer geometrical parameters (such as precise sample-to-detector distance and tilt of the detector) were optimized with respect to a NIST a LaB$_6$ (660a) standard.

The CeNi powdered sample was loaded in a membrane-driven diamond anvil cell fitted with a pair of 800~micron culet diamonds.
The gasket was made from 250~$\mu$m thick full-hard 301 stainless steel, pre-indented to about 120~$\mu$m and drilled with a 400~$\mu$m hole.
Gold was added to the sample as a pressure manometer and silicone oil was used as pressure transmitting medium.
The pressure on the DAC membrane was increased in 25 steps up to a maximum sample pressure of $\sim$7.8~GPa.
The pressure values were determined by fitting the measured gold unit cell volume to a third-order Birch-Murnaghan equation of state (EOS)\cite{Murnaghan,Birch} using the parameters V$_\mathrm{0} = 67.850$~\AA$^3$, B$_\mathrm{0T} = 167$~GPa, and B$^{\prime}_\mathrm{0T} = 5.77$, where V$_\mathrm{0}$ is the Au primitive cell volume at ambient conditions, B$_\mathrm{0T}$ is the bulk modulus, and B$^{\prime}_\mathrm{0T}$ is its first pressure derivative.\cite{Fei}

Time-of-flight neutron diffraction measurements at 100~K were performed at the Spallation Neutrons and Pressure (SNAP) beamline of the Spallation Neutron Source at Oak Ridge National Laboratory. Neutron detectors were located at 50~deg. and 90~deg. with respect to the incident beam. The sample-to-detector distance was 50~cm. The accessible $2\Theta$ range in this configuration extended from 30~deg. to 115~deg., up to ±7.5~deg. out of the scattering plane. The wavelength range of the neutrons was between 0.3~\AA and 3.7~\AA, which allows a sampling of Bragg reflections between $0.5 < d < 8$~\AA. The incident beam final collimation is a rectangular aperture measuring 1$\times$5~mm, designed to match the sample shape and the beam divergence. For sample loading a pellet was made from powdered CeNi mixed with Pb, which serves a dual role of pressure calibration standard and pressure transmitting medium. Pressure was generated in a Paris-Edinburgh high-pressure press fitted with boron nitride (BN) toroidal anvils\cite{Besson} and using a null scattering (TiZr) alloy that adds no coherent Bragg scattering to the data. The equatorial configuration of the cell was chosen allowing a wide angular aperture for the scattered beam.
For reference, we also measured the room temperature neutron diffraction pattern of the pressure cell filled with CeNi and Pb at SNAP and used the standard value of the Pb crystal lattice constant $a = 4.9496(3)$~\AA~\cite{Owen} for fitting.
To cool the pressure cell down to 100~K we used liquid nitrogen flow. After cooling, the Pb crystal lattice parameter was refined for every measurement to determine the actual pressure values using a third-order Birch-Murnaghan EOS\cite{Murnaghan,Birch} with measured value V$_\mathrm{0}$($T$=100~K)$ = 119.15$~\AA$^3$ and the parameters B$_\mathrm{0T}=48(5)$~GPa and B$^{\prime}_\mathrm{0T}=4(1)$ from work of Schulte and Holzapfel.\cite{Schulte}

To index the high-pressure (HP) phases we used the POWDERCELL\cite{Kraus} and DICVOL06\cite{Boultif} program packages. Neutron diffraction data reduction was done using the Mantid software suite.\cite{Arnold} Structural parameters were refined with the FullProf software.\cite{Carvajal}

\section{EXPERIMENTAL RESULTS}
\label{experiment}

\begin{figure}[tb!]
\begin{center}
\includegraphics[width=0.5\columnwidth]{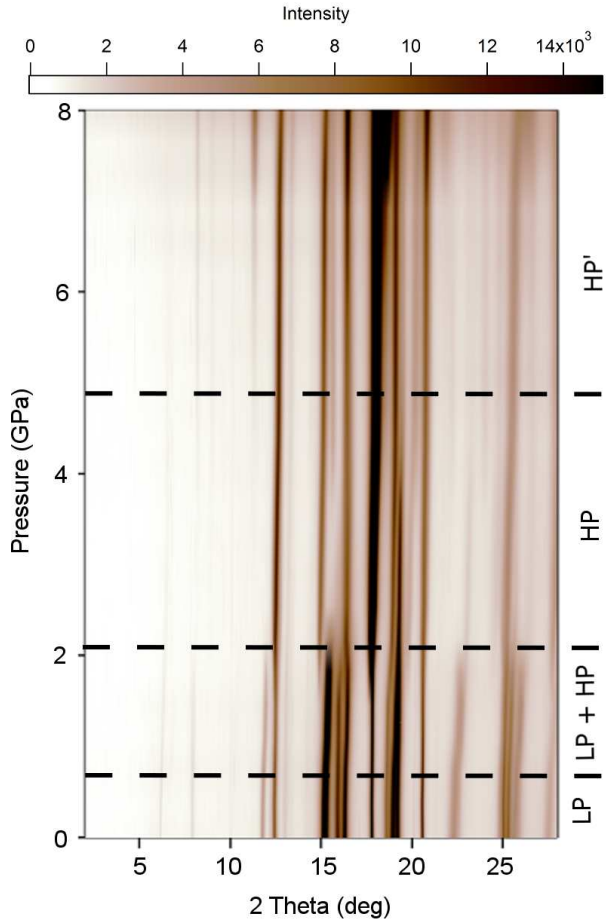}
\caption {(Color online) The contour plot of CeNi diffraction patterns as a function of pressure up to 8~GPa. Dashed lines separate the low-pressure (LP) CrB-structure, high-pressure (HP) FeB-structure, and HP$^{\prime}$ phase.\label{fig1}}
\end{center}
\end{figure}

Figure~\ref{fig1} shows the integrated X-ray diffraction (XRD) intensities for the CeNi sample along with the gold pressure standard as a function of increasing pressure at room temperature.
The peaks observed around 18 and 21 degrees are (111) and (200) Bragg reflections of Au used for pressure determination.
The low pressure (LP) CeNi structure remains stable from ambient pressure up to $P \sim 0.7$~GPa. Between 0.7 and 2.1~GPa the XRD patterns of CeNi exhibit remarkable changes. In particular, the appearance of new peaks around 15, 17.5 deg. and the suppression of diffraction peaks at ~8, 12, 15.5, 16 deg. clearly indicate a pressure-induced phase transition.
Above 2.1~GPa only subtle variations of peak positions and intensities are observed, however upon further increase in pressure a new diffraction pattern develops suggesting one more structural transformation starting at $P \sim 4.9$~GPa.

\begin{figure}[tb!]
\begin{center}
\includegraphics[width=0.5\columnwidth]{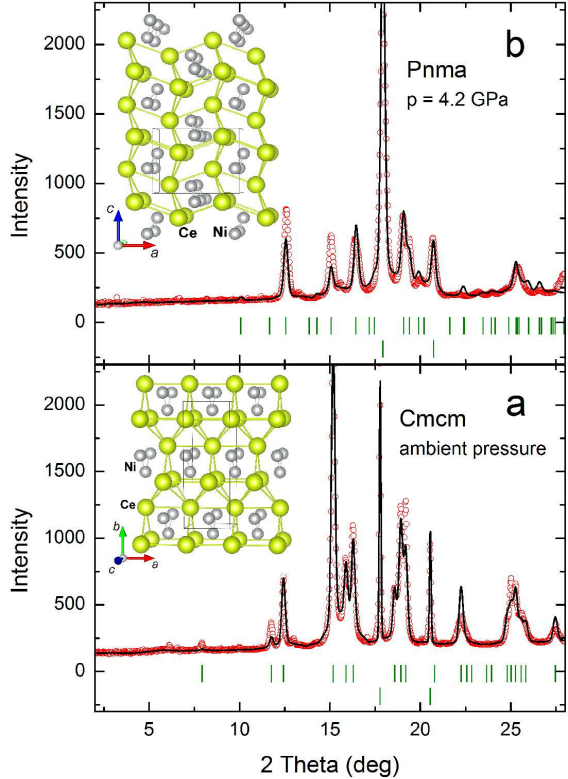}
\caption {(Color online) Powder diffraction data from CeNi at ambient pressure (a) and $P = 4.2$~GPa (b).
The red circles are the measured scattering intensity, and the black solid line represents the Rietveld refinement fit to the data.
The vertical bars indicate Bragg reflection positions of the main phase (top) and Au (bottom) used for pressure determination.
The inserts show schematic view of the low pressure CrB-type ($Cmcm$) (a) and high pressure FeB-type ($Pnma$) (b) of the CeNi crystal structure.\label{fig2}}
\end{center}
\end{figure}

Figure~\ref{fig2} shows typical XRD patterns obtained for the LP and HP phases of CeNi.
The CeNi LP phase (below $P = 0.7$~GPa) adopts the CrB-type orthorhombic structure described by the $Cmcm$ space group.
The CeNi primitive cell contains two Ce and two Ni ions, both located at (4c) $(0,y,1/4)$ Wyckoff positions.
A Rietveld refinement at ambient pressure resulted in the following structural parameters:
$a = 3.771(2)$~\AA, $b = 10.529(8)$~\AA, $c = 4.366(2)$~\AA, $y$(Ce)$ = 0.14(1)$, $y$(Ni)$ = 0.42(1)$, which is
in good agreement with values previously reported in the literature.\cite{Tinney,Gignoux}

The measured XRD pattern of the HP phase (pressure range between 2.1 and 4.9~GPa) was indexed to belong to the FeB-type orthorhombic crystal structure ($Pnma$ space group).
For example, the Rietveld refinement resulted in the following structural parameters at $P = 4.2$~GPa: $a = 7.161(5)$~\AA, $b = 4.390(4)$~\AA, $c = 5.086(4)$~\AA, $x$(Ce)$ = 0.13(1)$, $z$(Ce)$ = 0.20(1)$, $x$(Ni)$ = 0.09(1)$, $z$(Ni)$ = 0.66(1)$ (Fig.~\ref{fig2}).
Within the structural transition region at $P = 1.17$~GPa the X-ray diffraction pattern is described as a superposition of diffraction patterns resulting from 21$\pm$2 volume percent of CrB-phase and 79$\pm$7 volume percent of FeB-phase.
Attempts to refine the CeNi crystal structure from the powder diffraction patterns at pressures above 4.9~GPa (HP$\prime$ phase) did not enable us to extract reliable structural parameters or atomic positions.
Thus we cannot quantitatively describe the structure evolution in this pressure range.
It seems that above 4.9~GPa the diffraction patterns can no longer be ascribed to a single phase.
This issue will be discussed in Section~\ref{disc}.

\begin{figure}[tb!]
\begin{center}
\includegraphics[width=0.6\columnwidth]{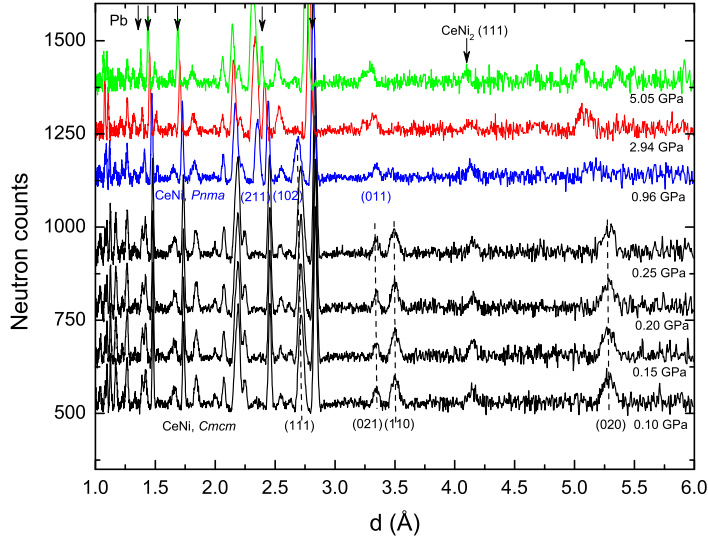}
\caption {(Color online) Pressure-dependent neutron powder diffraction patterns for CeNi obtained at SNAP at $T = 100$~K.\label{fig3}}
\end{center}
\end{figure}

\begin{figure}[tb!]
\begin{center}
\includegraphics[width=0.5\columnwidth]{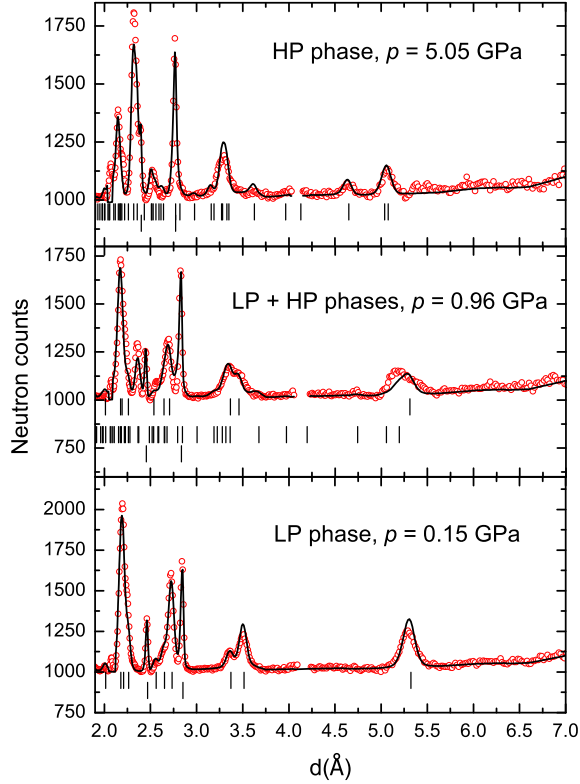}
\caption {(Color online) Observed (circles) and calculated neutron diffraction patterns of CeNi at pressure 0.15~GPa (LP CrB structure, bottom panel), 5.05~GPa (HP 3$\times a_{\mathrm{FeB}} Pnma$ structure, top panel), and 0.96~GPa (mixed LP-HP state, middle panel). Vertical bars indicate Bragg positions for LP phase (the upper rows at 0.15 and 0.96~GPa), for Pb (the lower row), and for HP phase (the middle row at 0.96 and the upper row at 5.05~GPa). \label{fig4}}
\end{center}
\end{figure}

In order to extend the phase diagram to lower temperatures we performed neutron powder diffraction study of the CeNi structure under pressures up to $P = 5.05$~GPa at $T = 100$~K.
Our experimental data are summarized in Fig.~\ref{fig3}.
It is seen that the initial CrB-type structure is conserved at low pressures.
For example, a FullProf refinement gives the following structural parameters at $P = 0.15$~GPa: $a = 3.704(2)$~\AA, $b = 10.566(5)$~\AA, $c = 4.342(2)$~\AA, $y$(Ce)$ = 0.135(2)$, $y$(Ni)$ = 0.422(1)$ (Fig.\ref{fig4}, bottom).
At $P = 0.96$~GPa a modification in the CeNi diffraction pattern is seen that is compatible with the expected $Cmcm \rightarrow Pnma$ symmetry change.
In fact, the $Cmcm$ (111), (110), and (020) reflections fully disappear between 0.96 and 2.94~GPa.
On the other hand, the representative $Pnma$ reflections, for instance, (211) and (011) appear in the same pressure region.
This means that CeNi transforms into a mixed low-pressure-high-pressure state.
At $P = 2.94$ and 5.05~GPa the diffraction pattern varies again as compared to that observed at the lower pressures.
In addition to the reflections of $Pnma$ FeB phase, new Bragg peaks appear, the most noticeable at $d \geqslant 5$~\AA.
Indexing the CeNi diffraction pattern at $P = 5.05$~GPa with DICVOL06\cite{Boultif} leads to a new orthorhombic cell with crystal lattice parameters close to those found for the quenched modification of the TbNi compound, $a = 21.09$~\AA, $b = 4.22$~\AA, $c = 5.45$~\AA~ ($Pnma$ space group).\cite{Lemaire}
This structure is closely related to the FeB type structure with transformation equations $a_{\mathrm{TbNi}}=3\times a_{\mathrm{FeB}}$;
$b_{\mathrm{TbNi}}=b_{\mathrm{FeB}}$; $c_{\mathrm{TbNi}}=c_{\mathrm{FeB}}$.
Assuming the CeNi high-pressure phase at 100~K to have this type of structure (we will call it 3$\times a_{\mathrm{FeB}} Pnma$ for a convenience), we performed Rietveld refinement of the experimental data at 5.05~GPa and obtained a rather good agreement between the experimental and calculated diffraction patterns (Fig.~\ref{fig4}, top). The resulting crystal lattice parameters are $a = 20.008(6)$~\AA, $b = 4.290(4)$~\AA, $c = 5.210(3)$~\AA, close to that for the quenched TbNi composition.\cite{Lemaire}
Moreover, we were able to describe the experimental data at $P = 0.96$~GPa, assuming this state to be a mixture of low-pressure $Cmcm$ and high-pressure 3$\times a_{\mathrm{FeB}} Pnma$ structures (Fig.~\ref{fig4}, middle).
According to the FullProf analysis, the volume fraction of low-pressure phase is 37$\pm$1\%, and the high-pressure phase occupies 63$\pm$1\% of the sample volume.
However, we note that a complete FullProf refinement of the high-pressure diffraction patterns is hardly possible due to large number of fitting parameters (for 3$\times a_{\mathrm{FeB}} Pnma$ structure there are three nonequivalent Ce ions and three nonequivalent Ni ions located at (4c) $(x,1/4,z)$ positions).
Moreover, the peak broadening at high pressure leads to significant overlap of reflections making the fit of atomic positions ambiguous.
Finally, CeNi ingots always possess strong texture (or preferred orientations) which cannot be averaged out completely by milling the ingots to powder.
As a result, a texture effect is seen clearly even in the X-ray diffraction pattern measured at ambient pressure (Fig.~\ref{fig2}).
Because of these circumstances, the calculated diffraction patterns for $P = 0.96$ and 5.05~GPa in Fig.~\ref{fig4} were obtained with the atomic coordinates equal to those reported for TbNi in ref.~\onlinecite{Lemaire} and fixed.
A similar procedure of fixing atomic positions was also applied for the FullProf analysis of the mixed-phase room-temperature X-ray diffraction pattern at $P = 1.17$~GPa.

\section{DISCUSSION OF EXPERIMENTAL RESULTS}
\label{disc}

\begin{figure}[tb!]
\begin{center}
\includegraphics[width=0.6\columnwidth]{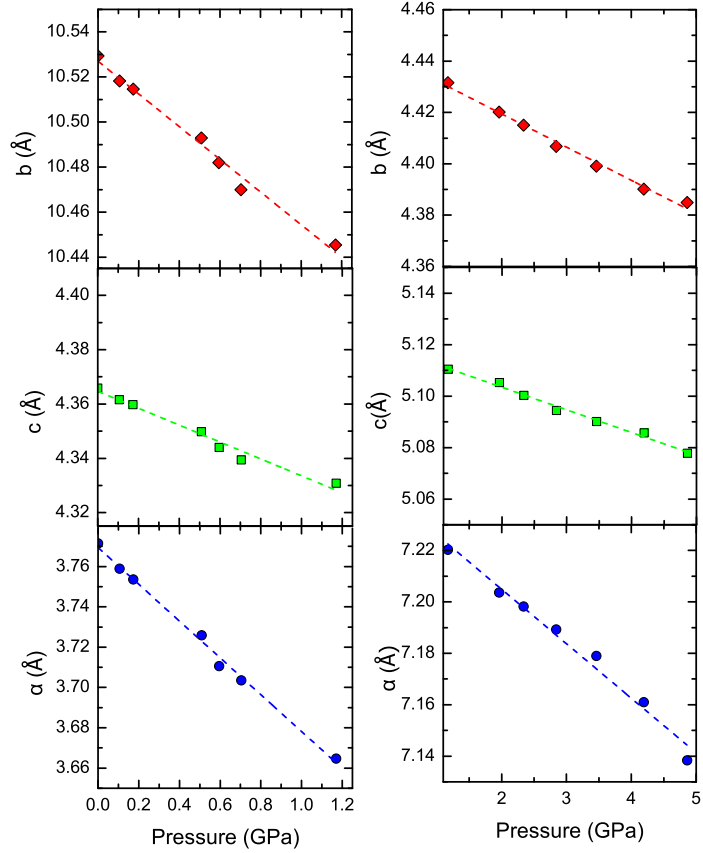}
\caption {(Color online) Pressure evolution of the crystal lattice parameters of low- (left) and high-pressure (right) phases of CeNi as obtained from the FullProf analysis of XRD diffraction patterns at room temperature. The solid lines are just a guide to the eye.  \label{fig5}}
\end{center}
\end{figure}

\begin{figure}[tb!]
\begin{center}
\includegraphics[width=0.8\columnwidth]{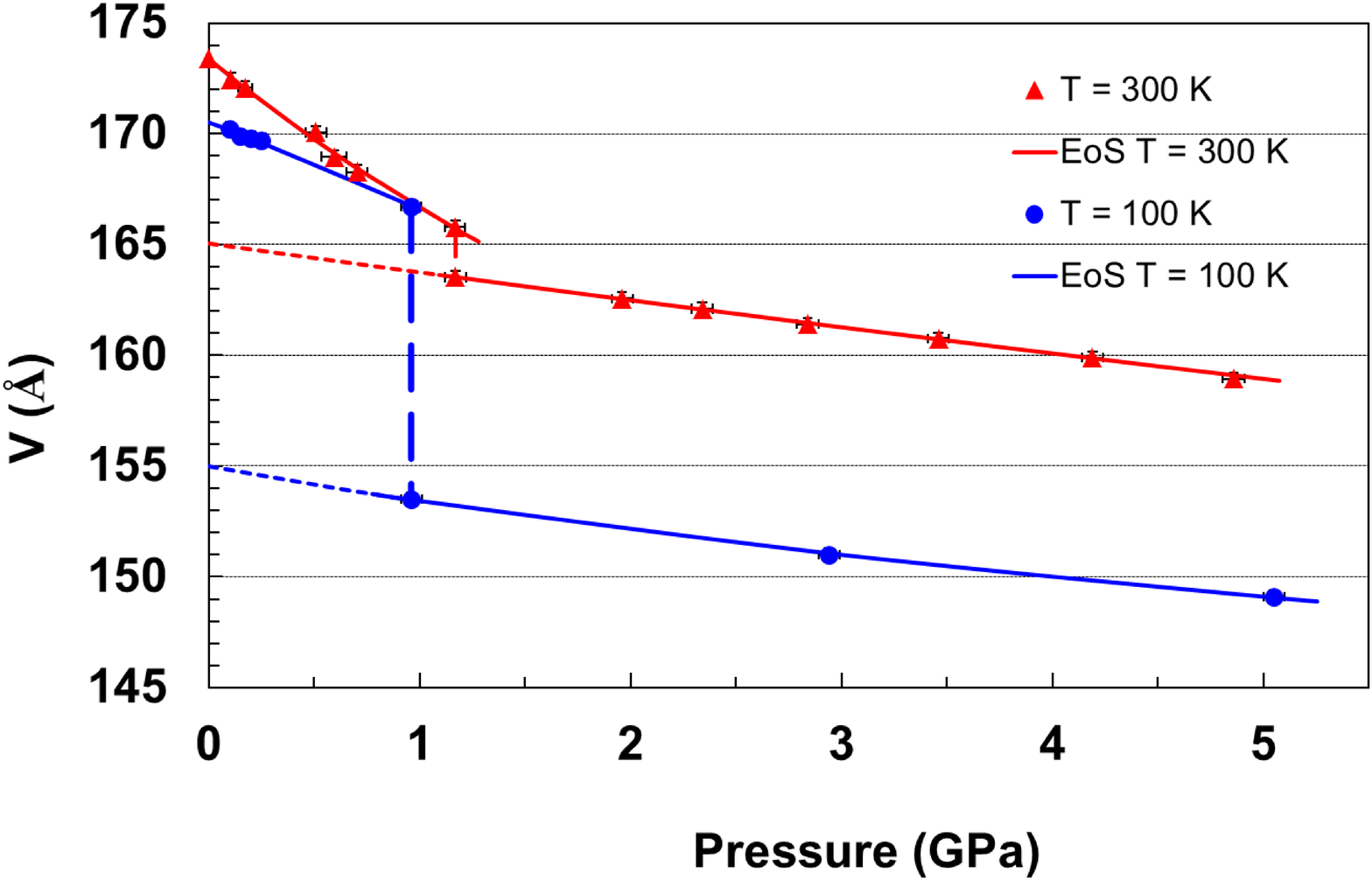}
\caption {(Color online) The unit cell volume of CeNi as a function of pressure as obtained from the XRD measurements at $T = 298$~K (red triangles) and from the NPD measurements at $T = 100$~K (blue circles). Solid lines are the results of fitting the measured unit cell volume to a third-order Birch-Murnaghan EOS\cite{Murnaghan,Birch} with the parameters given in Table~\ref{table1}. Note, that for the HP phase at 100~K the reduced to two formula units primitive cell $V = (a \times b \times c) /3$ is plotted.  \label{fig6}}
\end{center}
\end{figure}

Figure~\ref{fig5} shows the pressure dependences of the crystal lattice parameters for the LP CrB and HP FeB phases of CeNi as obtained from experimental XRD measurements at room temperature.
In the pressure range below 1.2~GPa (LP phase including phase coexistence region) the parameters $a$, $b$, and $c$ shrink by $\sim$2.8\%, $\sim$0.8\%, and $\sim$0.8\%, respectively, so that the change in volume is about 4.4\%.
The parameters $a$, $b$, and $c$ of the HP FeB lattice are reduced by $\sim$1.1\%, $\sim$1.1\%, and $\sim$0.6\%, respectively, when going from $P = 1.17$~GPa to 4.86~GPa.
The corresponding change in volume of the HP phase is about 2.8\%.
For a comparison, as seen from Fig.~\ref{fig6}, the volume jump at the transition ($P = 1.17$~GPa) is $\sim$1.3\% only at room temperature ($T = 298$~K).
However at lower temperature ($T = 100$~K, $P = 0.96$~GPa) the jump increases to 7.9\%.
Volume vs. pressure dependences shown in Fig.~\ref{fig6} were analyzed using a third order Birch-Murnaghan EOS.\cite{Murnaghan,Birch}
The obtained EOS parameters (see Table~\ref{table1}) indicate that the CeNi HP $Pnma$ structure is less compressible than the LP $Cmcm$ phase.

\begin{table*}[htb!]
\caption{\label{table1} Equation of state parameters for CeNi crystalline phases obtained by a fit of a third-order Birch-Murnaghan EOS\cite{Murnaghan,Birch} to the experimental X-ray ($T = 298$~K) and neutron ($T = 100$~K) diffraction data.
The same parameters resulting from DFT calculations are also given.
V$_\mathrm{0}$, B$_\mathrm{0T}$ and B$^{\prime}_\mathrm{0T}$ are the unit cell volume, bulk modulus, and its first pressure derivative at ambient pressure and at given temperature. Implied values B$^{\prime \prime}_\mathrm{0T}$ of the second pressure derivative of bulk modulus are also indicated.
Note, that for the HP phase at 100~K the reduced to two formula units primitive cell V = $(a \times b \times c)/3$ is given.}
\begin{ruledtabular}
\begin{tabular}{lcccccc}
&\multicolumn{2}{c}{$T=298$~K}&\multicolumn{2}{c}{$T=100$~K}&\multicolumn{2}{c}{DFT calculated}\\
&LP phase&HP phase&LP phase&HP phase&LP phase&HP phase\\
\hline
V$_\mathrm{0}$ (\AA$^3$)&173.4(2)&165.0(1)&170.5(2)&155.0(2)&164.6&162.3\\
B$_\mathrm{0T}$ (GPa) (\AA$^3$)&22(3)&124(11)&50(17)&88(4)&57&164\\
B$^{\prime}_\mathrm{0T}$&7(6)&3(4)&-12(22)&22.4(1)&18&20\\
B$^{\prime \prime}_\mathrm{0T}$&-1.0&-0.03&-5.5&-4.1&$-$&$-$\\
\end{tabular}
\end{ruledtabular}
\end{table*}

\begin{figure}[tb!]
\begin{center}
\includegraphics[width=0.8\columnwidth]{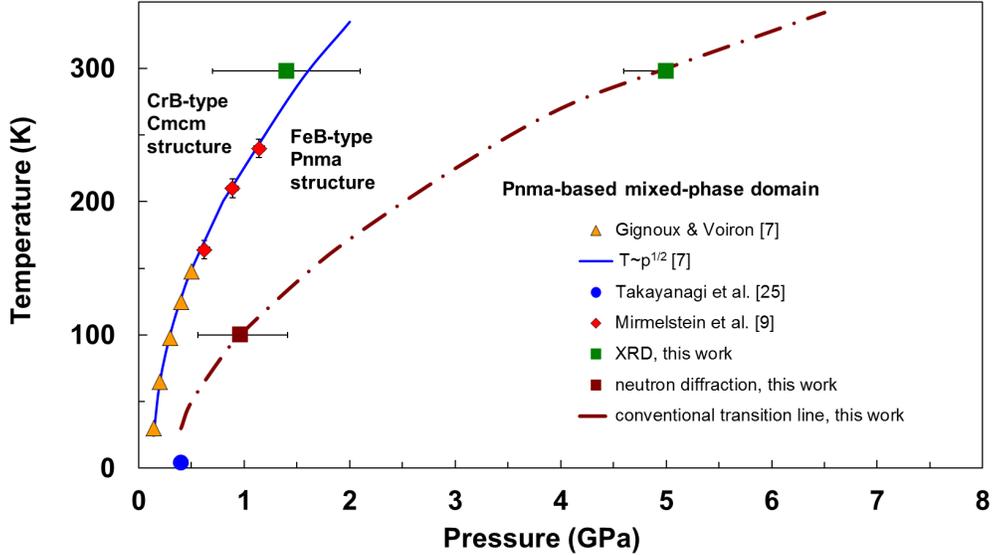}
\caption {(Color online) Approximate $P-T$ phase diagram of CeNi as follows from the X-ray and neutron powder diffraction measurements. Green squares correspond to the room-temperature transitions from CrB ($Cmcm$) low-pressure to FeB ($Pnma$) high-pressure phase and the second transition to the $Pnma$-based mixed phase region (HP$^{\prime}$ phase in Fig.~\ref{fig1}).
Yellow triangles and blue solid line represent the experimental data (magnetic measurements) and their approximation, respectively, from Ref.~\onlinecite{Gignoux}. The results of magnetic measurements from Ref.~\onlinecite{Mirmelstein} and specific heat data from Ref.~\onlinecite{Takaynagi} are shown by red diamonds and blue circle, respectively. Dashed-dotted line indicates the conventional transition line from the FeB-type of structure to the $Pnma$-based mixed phase state. \label{fig7}}
\end{center}
\end{figure}

The experimental results described above allow us to draw some conclusions about the $P-T$ phase diagram of CeNi (Fig.~\ref{fig7}).
First, we found that at room temperature the LP $Cmcm$ structure of CeNi transforms into the FeB-type of structure belonging to the $Pnma$ space group.
The latter structure is typical for the $R$Ni compounds where $R$ is the rare-earth metal from the second half of the lanthanide series, while light lanthanides, including cerium, form the crystal lattice of the CrB type.\cite{Dwight}
Therefore, the $Pnma$ symmetry is more favorable for the small rare earth ionic volume than $Cmcm$.
Both the FeB and CrB type structures contain a common structural unit, the trigonal prism, which is stacked differently to form either structure.\cite{Lemaire,Dwight}
If the $z$ value of the $4c$ sites in the $Pnma$ structure goes to zero, one obtains the higher-symmetry $Cmcm$ structure.\cite{Heatman}
Thus, the pressure-induced conversion from a CrB to a FeB type structure is not surprising. Furthermore, Hohnke and Parthé\cite{Hohnke} established the set of transformation equations between the crystal lattice parameters of these two structures:
\begin{eqnarray}
a_{\mathrm{FeB}} = \frac{2a_{\mathrm{CrB}} \times b_{\mathrm{CrB}}}{(a^2_{\mathrm{CrB}} + b^2_{\mathrm{CrB}})^{1/2}}; \quad  b_{\mathrm{FeB}} = b_{\mathrm{CrB}}; \quad
c_{\mathrm{FeB}} = \frac{(a^2_{\mathrm{CrB}} + b^2_{\mathrm{CrB}})^{1/2}}{2}
\label{eq1}
\end{eqnarray}

For this “ideal” case the volumes of FeB and CrB structures are equal. The room temperature lattice parameters of the CrB structure at $P = 1.17$~GPa (two phases coexist at this pressure) are $a = 3.665$~\AA, $b = 10.446$~\AA, $c = 4.331$~\AA (Fig.~\ref{fig5}).
For these parameters, Eqs.~(\ref{eq1}) predict the FeB structure to have $a = 6.92$~\AA, $b = 4.33$~\AA, $c = 5.53$~\AA, so that deviations from the experimental values at $T = 300$~K and $P = 1.17$~GPa are -4\%, -2\%, and +7\%, respectively.
The fact that the experimental FeB crystal lattice parameters are found to be close to the expected “ideal” values is an additional argument in favor of $Pnma$ symmetry of room temperature CeNi phase with pressure range between 1 and 5~GPa. The deviation of the experimental unit cell parameters from the “ideal” values is the result of a $\sim$1.3\% volume jump due to the transition.
Similar behavior of the unit cell parameters is found also at $T = 100$~K.
At $P = 0.96$~GPa we have $a = 3.643$~\AA, $b = 10.545$~\AA, $c = 4.340$~\AA for LP phase.
Equations (\ref{eq1}) give $a = 6.89$~\AA, $b = 4.34$~\AA, $c = 5.58$~\AA, while experimental values are $a/3 = 6.693$~\AA, $b = 4.297$~\AA, $c = 5.337$~\AA.
The volume jump at the transition ($T = 100$~K, $P = 0.96$~GPa) is $\sim$7.9\%, that is, as expected, much bigger than at room temperature.

Now we can establish an approximate $P-T$ phase diagram of CeNi (Fig.~\ref{fig7}).
At $T = 298$~K, the $Cmcm \rightarrow Pnma$ (CrB $\rightarrow$ FeB) transformation starts at $\sim$0.7~GPa and ends at $\sim$2.1~GPa (see Fig.~\ref{fig1}).
The middle point $P = 1.4$~GPa almost coincides with the empirical $P \sim T^2$ transition line reported in Ref.~\onlinecite{Gignoux}.
At $T = 100$~K and $P = 0.96$~GPa the CeNi sample is found to be inside the transition region (Fig.~\ref{fig6}), however not with the pure FeB $Pnma$ structure but its 3$\times a_{\mathrm{FeB}} Pnma$ modification.
We did not observe the pure FeB structure at 100~K.
Thus, we cannot say whether or not there is a pressure window for the FeB structure between the transition line established in Ref.~\onlinecite{Gignoux} and confirmed in Ref.~\onlinecite{Mirmelstein} by magnetic measurements, and the transition to the 3$\times a_{\mathrm{FeB}} Pnma$ at around 1~GPa (see Fig.~\ref{fig7}).
We believe that the FeB structure \emph{may} appear here and then transforms, almost immediately, under increase in pressure into the observed 3$\times a_{\mathrm{FeB}} Pnma$ structure.
Such a behavior cannot be excluded since one of these structures (CrB and FeB) can transform to the other directly or via a sequence of polymorphous transformations.\cite{Lemaire,Hohnke}
In fact, a very similar picture is observed at 300~K.
The CrB $\rightarrow$ FeB conversion does occur at 300~K, and then a second structure modification takes place at pressure above 4.9~GPa. 
As mentioned in Section~\ref{experiment}, the X-ray diffraction patterns of CeNi above this pressure can no longer be described by a single phase, neither pure FeB nor its polymorphous modifications of the $n \times a_{\mathrm{FeB}} Pnma$ type ($n = 2,3$ etc.).
Therefore, it seems quite reasonable to assume that at $P > 4.9$~GPa and at room temperature the CeNi sample consists of a mixture of HP FeB phase and $n \times a_{\mathrm{FeB}} Pnma$ structures, which can be called “$Pnma$-based mixed phase structure”.
If this is true, then the approximate CeNi $P-T$ phase diagram takes the form shown in Fig.~\ref{fig7} with the second transition line separating the FeB and $Pnma$-based mixed phase structural states.
Note, that an extrapolation of this transition line to $T = 0$~K goes to a critical pressure value $P \sim 0.4$~GPa, close to that obtained by specific heat measurements under pressure in Ref.~\onlinecite{Takaynagi}.
We cannot exclude, however, a more complicated CeNi structure at pressures above, say, 5~GPa.
The appearance of low scattering angle Bragg reflections in the CeNi XRD pattern at room temperature can also be considered to be the result of a coexistence of mixed FeB-CrB stacking variants similar to that found in $R_{1-x}R^{\prime}_x$Ni systems ($R =$~Gd and $R^{\prime} = $~Tb).\cite{Klepp}
However, the presence of the FeB-CrB mixture at high pressures does not look plausible since it requires reappearance of CrB phase.
The question remains open whether such a coexistence of phases is of intrinsic origin or is a consequence of non-hydrostatic compression conditions and/or stress between grains of the powder samples.
Nevertheless, based on our experimental results we can conclude that the structure of pressure-induced CeNi phase has $Pnma$ symmetry at least at pressures below $\sim$5~GPa.

The relative stability of the $Cmcm$ and $Pnma$ structures of CeNi under pressure can be understood by using density functional theory (DFT). Note, that we restrict our theoretical analysis to the CrB ($Cmcm$) $\rightarrow$ FeB ($Pnma$) transition.
First-principles calculations were performed using the Vienna \textit{ab initio} simulation package (VASP)\cite{Kresse1} based on projector augmented wave (PAW) pseudopotentials\cite{Kresse2} within the generalized gradient approximation (GGA) as parameterized by Perdew, Burke, and Ernzerhof (PBE).\cite{Perdew}
After careful convergence tests, a plane wave cutoff energy of 350~eV and 15$\times$5$\times$11 $k$-point Monkhorst-Pack mesh\cite{Monkhorst} for the 8-atomic unit cell of CeNi $Cmcm$ LP phase and 9$\times$13$\times$11 mesh for the $Pnma$ HP phase were found to be sufficient to converge the total energy within 2~meV/cell.
Bulk moduli have been computed by fitting to a Birch-Murnaghan equation of state.\cite{Murnaghan,Birch}

\begin{figure}[tb!]
\begin{center}
\includegraphics[width=0.8\columnwidth]{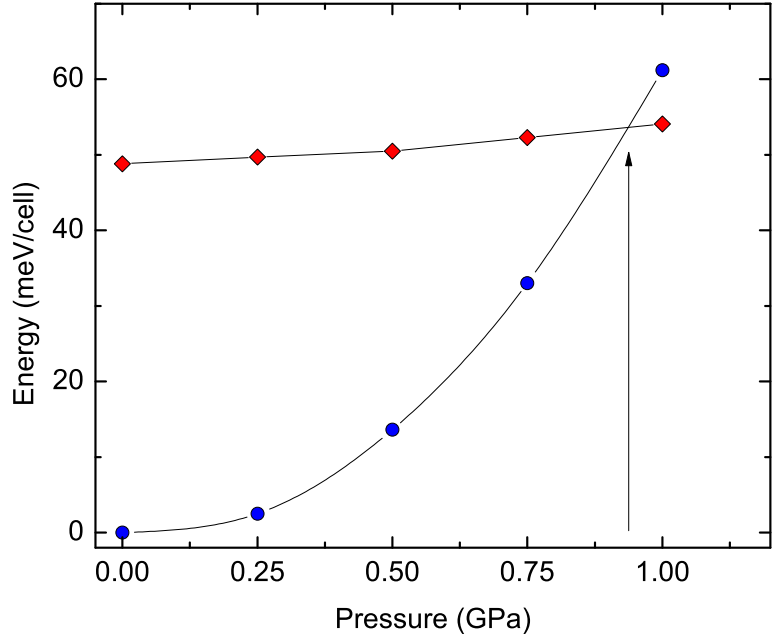}
\caption {(Color online) The relative energies as function of pressure for LP $Cmcm$ (circles) and HP $Pnma$ (diamonds) structure as obtained from DFT calculations. Arrow indicates the structural transition at pressure $P = 0.94$~GPa. \label{fig8}}
\end{center}
\end{figure}

First, we started from the ground states of the $Cmcm$ and $Pnma$ structures at ambient pressure.
We determine an equilibrium state by starting from experimentally determined parameters and letting the lattice constants and the atomic positions relax.
After the systems have relaxed the total energy difference between them turns out to be only 0.049(2)~eV/cell, with the $Cmcm$ structure being the lower one.
Next, we carried out calculations of the total energy of CeNi in both phases for different lattice parameters, which can imitate the influence of pressure.
For every pressure the atomic positions were allowed to relax keeping the lattice parameters fixed.
One sees in Fig.~\ref{fig8} that with increasing pressure (or with decrease in the lattice parameters in accordance to that found in our XRD measurements) the energies of the LP and HP phases cross at about 0.94~GPa.
This confirms the stability of the $Cmcm$ CeNi crystal structure at ambient pressure down to the lowest temperature and, at the same time, provides an explanation for the $Cmcm \rightarrow Pnma$ structural phase transition at $\sim$1~GPa.
The computed unit cell volumes and bulk moduli are found to be in a reasonable agreement with the experimental values (Table~\ref{table1}). Note, that the calculated bulk modulus of the $Cmcm$ structure turns out to be lower than that of the $Pnma$ structure, consistent with the experiment.

\begin{figure}[tb!]
\begin{center}
\includegraphics[width=0.6\columnwidth]{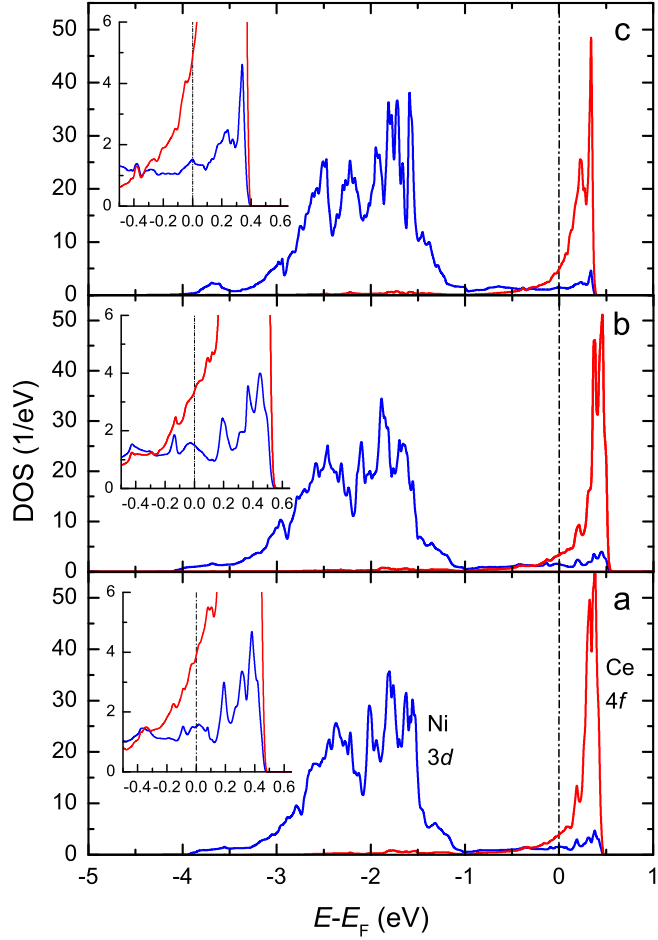}
\caption {(Color online) Site projected density of state for (a) CeNi $Cmcm$ phase at ambient pressure, (b) $Cmcm$ phase at $p=1$~GPa and (c) $Pnma$ phase at $p=1$~GPa (c). Zero in the horizontal axis represents the Fermi energy ($E_{\textrm{F}}$). The insets show the same DOS in the vicinity of $E_{\textrm{F}}$ on the expanded scale. \label{fig9}}
\end{center}
\end{figure}

The electronic densities of states (DOS) for CeNi on both sides of the structural phase transition resulting from our DFT-GGA calculations are shown in Fig.~\ref{fig9}(a,c).
For comparison, the anticipated DOS for a $Cmcm$ phase at $P = 1$~GPa is also shown (Fig.~\ref{fig9}(b)).
As expected, DFT calculations give metallic behavior of CeNi for both LP and HP phases, as indicated by a non-zero DOS at the Fermi level ($E_{\textrm{F}}$).
The site projected DOS of Ni 3$d$ and Ce 4$f$ orbitals overlap in the vicinity of $E_{\textrm{F}}$, suggesting their hybridization.
Fig.~\ref{fig9} demonstrates a rather weak but distinct impact of a moderate pressure, $\sim1$~GPa, on the electronic states of CeNi.
It can be seen that the structural transformation affects the DOS more than the pressure in the persistent structure, the Ni 3$d$ states near $E_{\textrm{F}}$ being the most affected.

As mentioned in Introduction, CeNi is a well known intermediate-valence system, characterized by the magnetic susceptibility of enhanced Pauli type at low temperature and passing through a broad maximum at around 140~K, a moderately high value of the electronic specific heat coefficient $\gamma \thickapprox 65$~mJ~mol$^{-1}$~K$^{-2}$,\cite{Gignoux3} and the characteristic spin fluctuation energy (Kondo scale) $T_{\textrm{K}} \thickapprox 30$~meV.\cite{Lazukov}
The intermediate-valence behavior of CeNi implies a rather strong hybridization of the 4$f$ electrons with Ni 3$d$ states.
The pressure-induced volume-collapse structural transition indicates essential changes of the CeNi electronic states.
What can be deduced about these changes from the structural data obtained in the present study?
Let us consider the variations of interatomic distances in CeNi due to the structural transition from the ambient pressure CrB-type of structure to FeB-type of structure ($P = 4.2$~GPa) at room temperature.
Ce ions in the $Cmcm$ structure have seven Ni ions in the first coordination sphere at nearly equal distances R$_1$(LP) = 2.938~\AA~ (2 ions), R$_2$(LP) = 2.948~\AA~ (4 ions), and R$_3$(LP) = 3.011~\AA~ (1 ion).
The next coordination sphere consists of eight Ce ions at R$_4$(LP) = 3.601~\AA~ (2 ions), R$_5$(LP) = 3.752~\AA~ (4 ions), and R$_6$(LP) = 3.771~\AA~ (2 ions).
In the $Pnma$ structure, Ce ions are surrounded by six nearest neighboring Ni ions at R$_1$(HP) = 2.357~\AA~ (1 ion), R$_2$(HP) = 2.761~\AA~ (1 ion), R$_3$(HP) = 2.794~\AA~ (2 ions), and R$_4$(HP) = 2.980~\AA~ (2 ions).
The second coordination sphere of the $Pnma$ structure, as well as $Cmcm$ structure, contains eight Ce ions at R$_5$(HP) = 3.525~\AA (2 ions), R$_6$(HP) = 3.611~\AA~ (2 ions), R$_7$(HP) = 3.773~\AA~ (4 ions).
Thus, Ce-Ni interatomic distances shrink essentially due to $Cmcm \rightarrow Pnma$ transition.
Ce-Ce interatomic distances decrease also, but their variation is less pronounced.
The decrease of the Ce-Ni distances suggests an enhanced Ce 4$f-$ Ni 3$d$ hybridization in the high pressure CeNi phase as compared to the ambient pressure $Cmcm$ structure.
As a result, one can expect an increase of the characteristic magnetic fluctuation energy $T_{\textrm{K}}$, and a correlated decrease of the electronic specific heat coefficient $\gamma$ and the low temperature value of the Pauli-like magnetic susceptibility (both these values are generally inversely proportional to $T_{\textrm{K}}$, which plays a role of an effective band width).
Obviously, these expectations agree with the experimental results.\cite{Gignoux1,Mirmelstein,Takaynagi}
In fact, at an applied pressure of about 0.5~GPa both the electronic specific heat coefficient $\gamma$ (see Fig.~3 in Ref.~[\onlinecite{Takaynagi}]) and the magnetic susceptibility $\chi$($T \thickapprox 30$~K) (red curve in Fig.~1 of Ref.~[\onlinecite{Mirmelstein}]) decrease by the same factor of about 1.4.
Note, that the smoothing of the peak structure of Ni 3$d$ DOS in the vicinity of $E_{\textrm{F}}$ (Fig.~\ref{fig9}(c)) may also be interpreted as a result of an enhanced 4$f-$ 3$d$ hybridization due to the structural transition.
However, to achieve more direct and complete understanding of the evolution of the 4$f$ electronic states across the volume-collapse transition (including the variation of $T_{\textrm{K}}$ and the 4$f$ occupation number as a function of pressure) inelastic neutron scattering experiments are required.

\section{CONCLUSIONS}

By means of X-ray and neutron powder diffraction measurements we have shown that pressure induces in CeNi structural phase transitions from a low-pressure CrB-type of structure ($Cmcm$ space group) to high-pressure phases belonging to the $Pnma$ space group.
The experimental results allow us to draw an approximate $P-T$ phase diagram of CeNi according to which CeNi undergoes two successive phase transitions within the $P-T$ domain $P < 8$~GPa, $T \leqslant 300$~K.
The first transition converts the LP structure into the FeB ($Pnma$) structure (which was not, however, observed experimentally at $T = 100$~K), while the second transition separates the FeB structure (HP phase) and $Pnma$-based mixed phase HP$^{\prime}$ state.

DFT calculations predict that at ambient pressure the total energies of the $Cmcm$ and $Pnma$ crystal structures are finely balanced, the latter energy being by 0.05~eV per unit cell higher than the former one.
In terms of a DFT approach the pressure-induced structural phase transition in CeNi can be understood as a result of the higher compressibility of LP phase as compared to the HP phase: the LP total energy rises rapidly with pressure leading to a phase transition at the estimated pressure of 0.94~GPa.

\begin{acknowledgments}
We are grateful to M.V.~Ryzhkov for enlightening discussions on the DFT calculations.
Research at Oak Ridge National Laboratory’s Spallation Neutron Source was supported by the Scientific User Facilities Division, Office of Basic Energy Sciences, a U.S. Department of Energy. Part of this work was supported by the Materials Sciences and Engineering Division, a U.S. Department of Energy, Basic Energy Sciences. This research used resources of the National Energy Research Scientific Computing Center, which is supported by the Office of Science of the U.S. Department of Energy under Contract No. DE-AC02-05CH11231.
Lawrence Livermore National Laboratory is operated by Lawrence Livermore National Security, LLC, for the U.S. Department of Energy, National Nuclear Security Administration under Contract DE-AC52-07NA27344.  JGT wishes to thank the Office of Advanced Simulation and Computing for support of his travel to Oak Ridge National Laboratory for experiments at the Spallation Neutron Source.
Work at RFNC-VNIITF was supported in part by Contract B601122 between LLNL and RFNC-VNIITF.
This research used resources of the Advanced Photon Source, a U.S. Department of Energy (DOE) Office of Science User Facility operated for the DOE Office of Science by Argonne National Laboratory under Contract No. DE-AC02-06CH11357.
\end{acknowledgments}

\bibliographystyle{apsrev}

\end{document}